February 11, 2019

***Chiral-Electromagnetic Gravitational Theory of Every 'Thing'***

***Evolving Gelfand-Dirac Hamilton-Riemann Quantum Cosmology***

Geoffrey F. Chew
*Theoretical Physics Group*
*Physics Division*
*Lawrence Berkeley National Laboratory*
*Berkeley, California 94720, U.S.A.*

**Summary**

The term 'quc' is shorthand here for a *cosmological--not* a physical--'quantum-universe constituent'. Although self-adjoint Hilbert-space angular-momentum and momentum operators unitarily generate quc rotations and spatial displacements, no *single*-quc is a *physical* 'thing'. Photons, 3 generations of fermions, plus massive vector and scalar bosons--*all* physicist-deemed 'elementary'--as well as dark matter, galaxies and black holes, within an *evolving* universe are, we propose, each a *temporary* 'quc family' that aggregates 8 Noether-conserved quc *attributes*. No 'thing' is *elementary*.

Here proposed is a thing-*devoid* 'von-Neumann big bang'--that established a huge but *finite* and *permanent* set of speed-*c chiral* qucs, with 'masses' $M\hbar/2\tau c^2$, where $M = 1, 2, ...M_{max}$ and $\tau$ is 'universe age'. ('Quc-mass', although not *physically*-meaningful, has the same *dimensionality* as any *thing*'s 'mass'.)

Quc-*aggregate* (thing) creations *and* disappearances--i.e., 'evolution'--are *symmetry*-governed by an 8-parameter 'centered-Lorentz' (CL) Lie group. CL's 6-parameter SL (2, c) Riemann-geometric 'exterior' of quc rotations and *hyperbolic* spatial displacements, has a conserved algebra of quc angular-momentum together with quc momentum *times* $\tau$. A 2-parameter *non*-geometric CL-*central* algebra comprises a pair of *discrete* coupled and conserved quc attributes--electric charge and 'chirality'. Chirality—*without* any physics counterpart--renders CL a *supersymmetry* and provides our present universe with *nuclear* forces--'strong interactions'.

Aggregate evolution, with a *fixed* and *finite* set of qucs, proceeds via a Schrödinger equation whose CL-invariant self-adjoint *non*-diagonalizable *hamiltonian* has potential-energies dependent on *correlated* quc electric-charges and 'masses'. Hamiltonian *expectation* specifies *total* universe energy. Gravitational and electromagnetic quc-*pair retarded*-potentials have (we propose) led *without* 'uncertainty', from a purely 'bachelor-quc' *big-bang* at $\tau_0 = G^{1/2}$ (if $\hbar = c = 1$) *starting*-age (~$10^{-43}$ sec), to the currently-evolving thing-assortment (that includes the reader).

Bachelor-quc population continues *presently* to *far*-exceed that within *things,* but the hamiltonian *momentarily* is binding certain *low kinetic-energy* quc subsets into 'families'-- e.g., chiral quc-*pair* vector bosons and semi-chiral quc-*pair* neutrinos. *Physical* masses, for 3 generations of quc-*trio* charged leptons and quarks, relate to 'Higgs-cores' that each *pairs* qucs of *zero* chirality but *nonzero opposite*-sign electric charges--of *equal* 3-valued *magnitudes*.

**Introduction**

Hamiltonian-evolving *Gelfand-Dirac-Riemann* big-bang *quantum cosmology* differs profoundly from Einstein's 'general relativity', whose gravity emphasis ignored both electromagnetism and *geometrical* quantum-universe *evolution* from a *beginning*. Pursuing Milne's thinking rather than Einstein's, we here spatio-temporally locate our universe *inside* a *forward* Lorentz-Minkowski light-cone. The latter (cosmological, *not* physical) notion--a geometrically-curved, hyperbolically-unbounded universe--associates to a remarkable Lie group that first received physicist attention in Faraday-Maxwell (classical) electromagnetic theory.

'Special-relativistic' quantum-*field*-theoretic (QFT) *physics*-association of 'boosts' to this group, as well as to its SL (2, c) semi-simple counterpart, employs (Feynman) 4-vector *momenta--not* 4-vector *Dirac-coordinates.* Cosmological Riemann-geometric (quc--*not* field) 'boostable 4-vectors' comprise a set of Dirac's coordinates--*not* of his momenta.

This paper recognizes two *different* meanings for the term 'momentum'. *Dirac* meaning (*hamiltonian*-accompanying) is the Hilbert-space self-adjoint 'canonical-conjugate' of a 'Dirac coordinate'; *Riemann* meaning is an algebra member of a *geometrical* Lie group. For the *physics*-employed (6-parameter) *Euclid-geometrical* group the two meanings coincide; for the *here*-employed (6-parameter) *geometrical* SL (2, c), they do *not*--due to Riemannian *curvature* of Milne's *cosmologically*-geometrical unbounded *hyperbolic* 'base' 3-space.

A Lie-group *contraction* as universe age passes to *infinity* relates cosmological (Milne-Riemann) to physical (Euclidian) *geometrical* groups. But *present* universe age, although huge, is *finite*--our universe's present character being, still, *far* from that of its distant future.

'Age' of any spacetime location--*not* Hubble's photon-redshift-based age, although roughly-relatable thereto via the local-*rapidity* detected in cosmic background radiation (CBR)--is *here* (following Milne's thinking) *defined* to be the SL (2, c)-invariant 'Minkowski distance'--*not* a *geometric* distance--from the light-cone vertex. *Common*-age *spatial* locations occupy a hyperbolic 'base' 3-space with *curved* Riemann geodesics. This curvature is proportional to the *inverse* of (positive, cosmological) universe age—which here will be denoted by the symbol $\tau$.

According to this paper, universe-*birth*-age, $\tau_0$, was $G^{1/2}$ (roughly $10^{-43}$ sec). In the units here-employed, where $c = \hbar = 1$, $G$ is Newton's gravitational constant. (Late in this paper, units will be employed where *G also* equals 1.)

Uncovered by CBR measurement (at current age) has been a $g^2$ ('fine-structure constant')-order of magnitude *rapidity* for Mach-emphasized angular-momentum 'clumps' (that are huge at earth-scale). Mach never addressed $g^2$ issues, which involve cosmology with *Hilbert space*. The present paper's hyperbolic 3-space accommodates (retarded) *classical* gravitational and electromagnetic fields, [1] but a 'hamiltonian *quantum* universe' (hqu) might seem impossible. Problematic has been *absence* of *finite*-dimensional *unitary* SL (2, c) Hilbert-space representation—an absence that perplexed and frustrated Dirac--functioning as a *physicist.* [3]

In Ref. (1) the present author has proposed hqu feasibility through a Gell-Mann-evocative *electro-extension* of the unitary *unboundedly-dimensional* Hilbert-space SL (2, c)-representation uncovered in 1946 by (mathematician) Gelfand and described (two decades later) in a book by Naimark. [2] A *finite* and *fixed* (although huge) set of SL (2, c)-governed 'quantum-universe constituents' —*cosmological* (*not* physical) entities—*we* have dubbed 'qucs'. Quantum *fields* enjoy *no* hqu (cosmological) status.

We propose that hqu evolution, which includes *recent* (Darwin) evolution of earth-located *conscious* matter, has occurred through perpetual rearrange-ability of a huge although finite quc-set by 'force'-generating, retarded electro and gravitational *inter-quc* potentials within a non-diagonalizable self-adjoint *hamiltonian* that is invariant under an *8-parameter* 'centered-Lorentz' (CL) Lie group.

CL's 6-element SL (2, c) Riemann-geometric 'exterior' combines quc rotations with (fixed-$\tau$) *curved* (hyperbolic) 3-space quc displacements—a semi-simple analog of the *physics*-foundational, *flat*-3-space, 6-element Euclidean group.

CL's 2-element *center* comprises *non*-geometric 'electro-chiral' *compact* quantum-theoretic quc-displacements. Chirality--*without* physics status--was exposed, although *ignored,* by Gelfand. [2] *Certain* chiral-electro, quark-related, *forces* may be described as 'nuclear'. Quc-*mass*-generated forces have, within physics history, been called 'gravitational'.

Mathematical definition will here be given for *chirality*--the name we have chosen for the remarkable *quantum*-cosmological notion uncovered (but ignored) by Gelfand in his 'regular' *infinite*-dimensional Hilbert-space *unitary*-representation of SL(2,c)--that may be seen as employing (hamiltonian-accompanying) *Dirac-coordinates*. [2]

The here-proposed CL-invariant hamiltonian adds *single*-quc kinetic energies (that ignore both chirality *and* electric charge) to *retarded* quc-*pair* gravitational *and* electromagnetic potential-energies. 'Super-symmetrically' (via chirality), the hqu 'state-vector' has evolved by a Schrödinger equation, as universe-age $\tau$ has advanced from the positive Planck-scale 'big-bang' *starting*-age, $\tau_0 = G^{1/2}$ (with $c = \hbar = 1$). Thereby, at *any* age greater than or equal to $\tau_0$, *total* universe-energy equals hamiltonian-expectation.

The present paper attends to a *correlation* with universe *start* of the hamiltonian (inter-quc, retarded) *gravitational* potentials. Involved is a finite, although huge, set of von-Neumann phase-space [4] 'quc-masses'--the quotation marks here emphasizing *c-magnitude* for *any* quc velocity. *Direction* of the latter is specified by a *pair* of Dirac quc-*coordinates*.

A *large*-$\tau$ hqu-*approximate*--because gravity-*ignoring*--'conformal' symmetry allows a renormalizable *physical* QFT to be based on (hqu-*absent)* 'boost-able' (Feynman) energy-momentum 4-vectors. This *physics* energy-momentum 'boost' notion *lacks* counterpart in *Riemann-geometrically*-gravitational, chirally-electromagnetic, Hamilton-dynamical, Gelfand-Dirac (GD) quantum cosmology.

'Measurement' is a vague notion that many 'Copenhagen' physicists *attempted* to render quantum-theoretically foundational. *Here* the author sides with Einstein, Schrödinger and Gell-Mann (disregarding Bohr, Born, Bell, Dirac, Fermi, Oppenheimer, Pauli, Heisenberg, Schwinger, Tomonaga and many distinguished others). *Ignoring* measurement, Ref. (1) defines 'hqu reality' (for $\tau > \tau_0$) via universe-ray *expectations* of certain self-adjoint GD Hilbert-space operators--expectations that represent current-densities of electric charge and energy-momentum.

In defining *real* energy-momentum current density, Reference (1) *erroneously* associated 'quc energy' meaning to the symbol, *M*--*here* denoting a *positive-integer* proportional to discrete 'von Neumann quc mass'. (*Odd-M* values accompany *negative* quc electric charge while *even M* accompany *positive*. The *M*-integer of an electrically-*neutral* quc may be *either* even *or* odd.)

*Spanning* a real line is the *continuous* spectrum of the self-adjoint quc-*energy* operator--a *Dirac-momentum* that does *not* represent an (additive) Noether-conserved quc attribute. Various other correctable errors in Reference (1) are less foundational.

It is pedagogically unfortunate that the *mathematical* term, 'expectation', carries in 'ordinary' language a *probabilistic* meaning. Confusion has been increased by 'measurement' association with (philosopher-recognized, Darwin-evolution-related) *conscious*-thing 'free will'.

In the following section, the *magnitude* of Noether-conserved electric-charge distinguishes 'dark' qucs *and* 'baryonic' qucs from qucs that are 'bright'. Bridging of 'quantum-classical gap' by the *bright* (although tiny) quc charges within photons and electrons underlay humanity's glorious (*physical*) twentieth-century *atomic-molecular* science.

Long-before human physicists and physics, early-universe collaboration between gravity and chiral-electromagnetism we believe *created* dark, bright and baryonic energetic quc-aggregates--*no* aggregates having been present at a von-Neumann-gaussian 'quantum phase-space' [(4)] 'purely-*bachelor*-quc' hqu-beginning— a start we believe to be *mathematically* specifiable. The present paper offers a math-candidate for 'big-bang commencement'.

Noether-conserved by the hqu hamiltonian are electric charge, angular-momentum and Riemann-momentum times age—*7* aggregate-able quc-attributes with names familiar in human languages. The name 'chirality', as *here* employed, is *not* 'familiar'. (In the past, this word has been assigned a variety of meanings *different* from that employed here.) Aggregates comprise all 'things', including photons, leptons, quarks and dark matter, while *not* including a bachelor-quc 'reservoir'—presently still a quc-*majority*, with high kinetic energies that resist aggregation.

Hqu-*aggregate* meaning for 'matter' accompanies a *central* non-geometric CL-*algebra*--of electric charge and chirality--that *joins* geometric (exterior) Riemann-momentum (times $\tau$) and (Wigner) angular momentum. Quc-aggregates ('things') *additively* carry 8 Noether-conserved quc attributes. The latter differ from (*non*-conserved) aggregate attributes such as size and shape (e.g., 'photon double-helix')--features *un*-attributable to any *single* quc.

A discrete von-Neumann quc-attribute, 'quc-mass' (*non*-Noether, *not* additive), enjoys an extraordinary hqu status that *demotes*--to large-$\tau$, almost-flat 3-space, group-contraction *approximations*--such QFT concepts as energy *conservation*, particle mass and quark color.

Dirac quantum theory--employing 'complete' sets of commuting self-adjoint Hilbert-space operators (csco's)-- [(3)] *and* Gelfand's *unitary* Hilbert-space representation of SL (2, c)-- [(2)] have in Ref. (1) led the author to recognize the conserved CL-*central non-geometric* quc attribute, 'chirality', a GD-momentum whose dimensionality--the *same* as that of (geometric) angular momentum--differs from (and is independent of) the dimensionalities of electric charge and energy.

Notions of electric charge, of energy and of momentum and angular momentum have, in physics history, enjoyed continuous *and* discrete, classical *and* quantum, Noether-*conserved* status. Chirality is a *non*-geometric, purely-discrete, purely-quantum, *conserved* hqu feature that facilitates large-age *approximate* physics notions of particle masses and nuclear forces ('strong interactions'). Surprising to the author has been (finite-age) cosmological *absence* of (Noether) *energy*-conservation.

The following section calls a quc 'dark' if its charge-integer $Q$ is 0 and 'bright' if $Q = \pm 3$. With $Q = \pm 1$ or $\pm 2$, perhaps confusingly, we call a quc 'baryonic'. A *pair* of *zero*-chirality while *oppositely*-charged (bright *or* baryonic) constituent-qucs gives, to each of 3 *different* '*massive*-elementary-fermion' generations, a 'Higgs core' —that facilitates an *approximate* (physics) meaning for 'charged-fermion mass'.

A *single* charged 'valence' quc, of nonzero mass and with chirality *either* +1 *or* –1, specifies a massive-fermion electric charge. Dirac-*superposition of* +1 *and* –1 valence-quc chiralities ('Dirac-doubling') specifies massive-fermion *rapidity*. In quarks, the (mean) masses of *baryonic* (*not* bright) valence-qucs serve 'color duty'--by locating either 'below', 'comparable to' or 'above' the quark's Higgs-core mass. [In a charged-*lepton*, the (mean) mass of the *bright* valence-quc is, we believe, *always smaller* than the lepton's (mean, Higgs) *core*-mass.]

We believe the *present*-universe content called 'GUT' by QFT is provided by 'chiral-electromagnetic' hqu 'nuclear forces', without need for (a 'foundational') *color*. The impressive successes of 'color physics' the author expects to be *exceeded* as understanding of quantum cosmology advances.

Electric-charge extension of Gelfand's Hilbert space allows gravity *plus* chirality-guided electromagnetism to be represented by quc-*pair* retarded-potentials within the hqu hamiltonian. CL-algebra Noether-associates *conservation* of 8 universe attributes, as evolution proceeds via a Schrödinger equation, to CL-invariance of a (non-diagonalizable) hamiltonian whose expectation *specifies* total universe-energy. (The physics notion of 'stationary state' is cosmologically untenable.)

*Definition* of baryon number via quc electric charge [(1)] maintains Gell-Mann's relation between a *quark*'s charge and its baryon-number. But *one* quc (with kinetic although *no* potential energy) is *not* an aggregate; *one* quark is (approximately) a *chiral-electromagnetically*-stabilized aggregate of 3 *different* qucs. [(1)] Von-Neumann big-bang established a permanent *finite* set of *discretely-different,* $2\tau c^2$-spaced, positive 'quc masses'. Speed-*c* qucs, of *common* charge and chirality but *unequal* 'masses', *differ* from each other (regardless of their continuous positive kinetic energies).

Hyperbolic (geometric) 3-space curvature varies *inversely* with $\tau$, being completely $\tau$-determined (*unrelated* to *energy*-density). Hqu-age $\tau$ *presently* is so large that Euclid's flat 3-space geometry (used in astronomy, 'non-relativistic' physics *and* QFT) provides an accurate basis for ('Popper', human) *science*. But, for *big-bang* quantum cosmology, the huge *curvature* of *early*-universe *non*-Euclidean hyperbolic (Riemann-geometric) 3-space was essential.

Although for human astronomy and physics (including 'black-hole' description) Euclidean geometry suffices, early-hqu 'bootstrap'-evolution of bright-quc-pair ('double-helix') photon-aggregates occurred when 3-space curvature was enormous compared to that currently prevailing. A 'theory of everything' requires attention to our universe's history at ages when the earliest photons were evolving—via *gravity* plus chiral-electromagnetism. Photon *evolution*--to its current foundational status in human-physics--depended on gravity as well as on chiral-electromagnetism.

Risking reader confusion with (*non*-Riemannian) 'general relativity', we apply the term 'inflation' to the early hqu era of enormous while rapidly-decreasing 3-space curvature--between big bang at a Planck-scale age $\tau_0 > 0$ with *no* aggregated matter, and ages by which photon-aggregates had begun to emerge. Absence of early-evolution dynamics has heretofore left mysterious how photons (and QFT's 'color') developed during inflation. Hamiltonian 'double-helix quc-bootstrap' photon evolution, we believe, bridged an 'inflation gap'.

QFT has impressively represented human-observed particulate matter as composed of a finite set of 'elementary' particles; but because Euclidean geometry and 'boost-able'(Feynman) energy-momentum 4-vectors are foundational for QFT, the latter cannot explain *how* (so-called) ‘elementary’ particulate-matter developed. Astronomical observation of *QFT-incomprehensible* non-particulate ‘dark matter’ has encouraged a widely-held opinion (evidently shared by the author) that *arbitrariness* of QFT’s large set of ‘elementary-particle’ masses rules QFT *out* as ‘final theory’.

We expect GD Riemannian-Hamiltonian quc dynamics (via chiral-electro *and* gravitational forces) to elucidate the origin not only of QFT’s photons but of its *massive* ‘elementary’ bosons and 3 ‘generations’ of fermions. Hqu representation of the latter is discussed in a Ref. (1) appendix whose content (including neutrinos) will here be extended. (*Absence* of *zero*-chirality qucs from *any* of the QFT 'elementary' *vector*-bosons deserves immediate attention.)

The present paper deals with *quantum*-theoretic aspects of hyperbolic (non-Euclidean) Riemannian geometry. Quc interpretation of Gelfand math allows revival of hamiltonian-based Dirac quantum theory. [(3)] Appendices of Ref. (1) suggest that ‘elemental’ 7-valued quc electric-charge and 3-valued chirality *together* have allowed evolution of an initially aggregate-free hqu into current *galactic*-scale dark matter plus *micro*-scale particles, *some* of the latter collectively composing *macro*-scale condensed matter. *Here* we refine specification of the hamiltonian *gravitational* potential energy.

We propose an *initial* occupation of quc Hilbert *phase-space*--by a von-Neumann big-bang bachelor-quc collection--that correlated *sign* of quc electric charge with quc *mass*. Subsequent quc ‘marriages’, via a *first-order* (Schrödinger) differential equation, have evolved a cosmological (*not* physical) hqu Hilbert vector--creating not only ‘elementary’ particles (such as photons, leptons and quarks) plus dark matter, but aggregates to which humans have yet to assign names. Perpetual 8-element supersymmetry we believe essential to our universe’s evolution of macro-scale ‘conscious’ condensed-matter aggregates such as this paper’s author and readers.

**CL Definition: *Central-Extension* of Gelfand’s *Unitary* SL(2,c)-Representation**

The *initial* ($\tau = \tau_0$) hqu Hilbert vector--a von-Neumann phase-space [(4)] 'thingless big-bang'--has evolved, we propose, via a $\tau$-dependent non-diagonalizable hamiltonian that is invariant under an 8-parameter Lie group--CL. The latter comprises a semi-simple 6-parameter SL (2, c) ‘Riemann-geometric exterior’--of quc rotations and *curved* 3-space displacements--*plus* a *non*-geometric 2-parameter 'chiral-electro' *center* (commuting with full CL). This paper’s present section *defines* (for *any* post-big-bang age) the Lie group we call 'CL'.

SL (2, c) lacks *finite*-dimensional *unitary* Hilbert-space representation. Gelfand [(2)] found a unitary ‘GD’ representation via an *unbounded* set of Hilbert vectors that *each* is a *normed* complex differentiable function of a Dirac 6-*coordinate* csco--4 non-compact *geometrical* quc coordinates accompanied by a non-compact ‘semi-geometric’ quc ‘local-time’, *plus* a *non*-geometric quc Dirac coordinate that is compact and canonically-conjugate to 'chirality'. (Naimark's book *fails* to recognize importance *either* for chirality *or* for the canonical-conjugate thereof.)

The foregoing Dirac-coordinate sextet amounts to a complex 2×2 unimodular matrix $\boldsymbol{a}$. Gelfand’s *unitary* ‘regular’ SL (2, c) representation multiplies each such csco-sextet-matrix, from the *right,* by *another* complex 6-parameter 2×2 unimodular matrix, $\Gamma^{-1}$. We shall here *define* an 8-parameter group CL via a *finite*, Gell-Mann-evocative, compact 'electro-extension' of Gelfand's 6-Dirac-coordinate Hilbert space--an extension allowing *chirality* to 'team-up' with electric charge so as to generate what physicists call 'nuclear' forces ('strong' interactions).

Hqu 'inhabits' a Hilbert-space *tensor-product* of *seven* GD 6-coordinate wave functions, $\boldsymbol{\Psi}_Q(\boldsymbol{a})$--associating to qucs each of whose electric charge (in later-specified units) is proportional to an integer $Q$ that takes *one* of 7 different possible values. According to Ref. (1), a quc 'charge integer' $Q$ may be 0, ±1, ±2, or ±3. This $Q$-septet represents dark matter as aggregates of $Q = 0$ qucs. Pairs and trios of *electrically-charged* qucs represent *all* QFT 'elementary' particles.

At *any* post-big-bang age $\tau$ a tensor-product, $\boldsymbol{\Psi}^{\tau}(\{\boldsymbol{a}_Q\})$, *completely* specifies hqu. The hamiltonian—Formula (13) of the present paper--*determines* the $\tau > \tau_0$ hqu 'state-vector', by the Schrödinger equation (14), from a *quc-defining* von Neumann phase-space *initial-state* Hilbert-vector (*at* $\tau = \tau_0$). Formulas (2) and (5) below address Hilbert-vector norm. The hamiltonian *potential* energy *sums* electromagnetic and gravitational quc-*pair* retarded potentials—specified by Formulas (11) and (12). Inter-quark 'nuclear' forces are 'chiral-electro'.

How does the 8-parameter group CL relate to the foregoing? *Exteriorly,* Gelfand's 'regular' SL (2, c) representation displaces *each* $\boldsymbol{a}_Q$ via *right* unimodular 2×2 matrix-$\boldsymbol{\Gamma}^{-1}$ multiplication (6 'exterior' Riemann-geometric parameters). A *central* hqu CL element is a 2-parameter *non*-geometric *chiral-electro* shift (commuting with *all* CL elements). The *two* central (CL *sub*-group) parameters, *one* of these a 'gauge' angle $\omega$, are *both* compact.

*Centrally*, CL *left*-multiplies $\boldsymbol{a}_Q$ by a 1-parameter unimodular *diagonal* 2×2 matrix, $\boldsymbol{\gamma}$. Under a (*fixed*-age) $\omega$, $\boldsymbol{\gamma}$, $\boldsymbol{\Gamma}$-specified 8-parameter CL-element (preserving Hilbert-vector norm), any fixed-$Q$ factor, $\boldsymbol{\Psi}(\boldsymbol{a}_Q)$, within a tensor product, changes to an equivalent factor according to

$$\boldsymbol{\Psi}(\boldsymbol{a}_Q) \rightarrow_{(\omega,\, \gamma,\, \Gamma)} e^{iQ\omega}\, \boldsymbol{\Psi}(\boldsymbol{\gamma} \boldsymbol{a}_Q \boldsymbol{\Gamma}^{-1}). \tag{1}$$

Essential is commutativity in (1) of the left ($\boldsymbol{\gamma}$, *non*-geometrically chiral) and right ($\boldsymbol{\Gamma}^{-1}$, geometric) multiplications of $\boldsymbol{a}_Q$. The 2-parameter CL center is coordinated by the 'gauge' angle $\omega$ *and* by the 1-parameter diagonal unimodular 2×2 matrix $\boldsymbol{\gamma}$—writeable as $exp\,(-\boldsymbol{\sigma}_3 \Delta)$ with *imaginary* $\Delta$ and with $\boldsymbol{\sigma}_3$ the *diagonal* Pauli hermitian traceless *2×2 matrix*. The electro-chiral group center is *non*-geometric, whereas CL 's Riemann-geometric exterior is isomorphic to SL (*2, c*).

The 8-member CL *algebra* comprises a *Riemann-geometric* Lorentz 6-vector, associable to quc momentum times age *plus* quc angular momentum (see the section below on a quc kinetic-energy), *supplemented* by a *central* chiral-electro *non-geometric* 2-vector.

**Remarks**

With respect both to qucs and to aggregates thereof, the *pair* of *compact* Noether-conserved *central*-algebra (*non*-geometric) '*quality*-space' components we describe as: 'chirality' in units $\hbar/2$ and *electric charge*, in $g$-proportional units later specified whose dimensionality (because of a factor $c^{1/2}$) is *inexpressible* through that of chirality and energy. What we call 'dimensionality space' is spanned by the 3 dimensionalities of energy, electric charge and that *either* of age *or* of angular-momentum. Chirality-dimensionality is the *same* as that of the latter. The *full* CL algebra comprises 8 (diagonalizable) self-adjoint Hilbert-space operators.

Ref. (1) explains how dimensionless *baryon number*, conserved although not a CL-algebra member, is specified (following Gell-Mann) via $Q = \pm 1, \pm 2$, by 4 components of our 7-fold electric-charge extension of Gelfand's Hilbert space. 'Three-dimensionality of quality-space' is displayed by *independence* between the dimensionalities of electric-charge, energy and that *shared* by angular-momentum *and* chirality.

*The mostly non-commuting (see later* section on hamiltonian kinetic energy) members of CL's *exterior* 6-element algebra invoke *only* the *dimensionalities* of momentum times age and angular momentum—2 special 'directions within quality 3-space'. Eight (diagonalizable) self-adjoint Hilbert-space 'quc-momentum' operators Noether-associate to conserved *single*-quc attributes that are *additively* transmittable to aggregates.

*Single*-quc energy, although definable and proportional to a Dirac momentum, is *not* additively 'Noether-conserved'. *Total* universe energy, represented by hamiltonian expectation, is *not* conserved. *Physics* energy-conservation Noether-associates to Wigner's (huge-age, particle-*physics*-meaningful, but gravity and big-bang-ignoring) 10-parameter 'Poincare group'--with its *Euclidean* geometry and 'conformal invariance'.

Although an *expectation* definition of 'reality' is provided by self-adjoint quc 4-vector Dirac-*coordinate* operators (*without* using the canonical-conjugate of chirality), [(1)] plus CL-invariant quc *masses*, there are *no* quc 4-vector *Dirac-momentum* operators. There *is* a self-adjoint quc-momentum-*direction* unirrep-operator-*pair*--discussed in the later section on quc kinetic energy. This remarkable pair associates neither with Riemann nor with Dirac—*only* with Gelfand.

The 3-dimensional 'quality space' spanned by CL's algebra is occupied not only by *conserved* momenta of qucs and quc aggregates but also by *non*-conserved aggregate attributes of *spatial* or *temporal* dimensionality--attributes that individual qucs are *incapable* of carrying. (The *present* age of any 'thing'--the reader, for example, or the author--although never sharply-definable is always *smaller* than $\tau_{present}$-$\tau_0$.)

Conservation of chirality and electric charge, together with conservation of momentum times $\tau$ and of angular-momentum, follow (Noether) from hqu-hamiltonian CL invariance. Mach's rotational universe symmetry is extended—not only dimensionally (3 to 8) but from 'classical' to 'quantum'. Chirality *absence* (in contrast to electric charge), from the Reference (1) set of *reality-defining* self-adjoint current-density operators, has obscured chirality conservation. [The foregoing assertion seems to the author to be justified, despite the errors in Reference (1).]

Hqu 'initial condition' implies not only zero total-universe angular momentum but zero total chirality. Hqu's 'hyperbolic' although *globally*-conserved 3-momentum times $\tau$ is, for any *individual* quc, not only continuous but *unbounded*—a feature essential to unitarity of Gelfand's Hilbert-space SL (2, c) representation. Expectation is $\tau$-independent for all 3 components of continuous *total* hqu momentum *times* $\tau$. These 3 expectations *vanish* for the big-bang Hilbert-vector--an illustrative possibility for which will be given by Formula (10).

**Dirac Quc-Coordinate Csco-Sextet (for each of 7 different values of quc electric charge)**

Any complex repeat-ably-differentiable wave-function, $\boldsymbol{\Psi}(\boldsymbol{a}_Q)$, *unitarily* representing (fixed-age) CL by Formula (1), is a Hilbert vector with the CL-invariant (finite) norm,

$$\int d\boldsymbol{a}_Q \left| \boldsymbol{\Psi}(\boldsymbol{a}_Q) \right|^2. \tag{2}$$

The 6-dimensional CL-invariant volume element (Haar measure) $d\boldsymbol{a}_Q$ is below expressed for each $Q$ [Formula (5)], through a trio, $s_Q$, $y_Q$, $z_Q$, of *complex* continuous coordinates--a Dirac-*coordinate* (*not* Dirac-momentum) csco sextet collectively equivalent to the matrix $\boldsymbol{a}_Q$.

Through some *special* [such as Formula (10)] $\boldsymbol{\Psi}(\boldsymbol{a}_Q)$ dependence on $Re\ s_Q$ --i.e., on each quc's fixed-age 'local time'--*at big bang*, a von Neumann-phase-space gives *discretized* meaning to a finite huge set of positive 'quc-masses'. Because the present section considers separately each *single (*fixed*)* value of the integer $Q$, within the remainder of this section we often *omit* the 7-valued electric-charge subscript. (The positive quc-*mass* integer *will* here below emerge.)

A (dimensionless) sextet of (commuting) GD quc *coordinates*, the csco, *Re s, Im s, Re y, Im y, Re z, Im z*, is definable through a *Pauli-matrix* formula for a *general* 2×2 unimodular matrix,

$$\boldsymbol{a} = exp\ (-\boldsymbol{\sigma}_3 s) \times exp\ (\boldsymbol{\sigma}_+ y) \times exp\ (\boldsymbol{\sigma}_- z), \tag{3}$$

each member of the *real*-matrix pair, $\boldsymbol{\sigma}_\pm \equiv \frac{1}{2}\ (\boldsymbol{\sigma}_1 \pm i\boldsymbol{\sigma}_2)$, squaring to a *zero* matrix. According to Formula (3), displacement by an *imaginary* increment, $\Delta$, of the (complex) coordinate $s$, at *fixed* $Re\ s$, $y$, $z$, follows (with exterior $\boldsymbol{\Gamma}$ fixed) when $\boldsymbol{a}$ is multiplied, from the *left*, by $exp\ (-\boldsymbol{\sigma}_3 \Delta)$.

Each of the 3 *anti*-commuting Pauli matrices, $\boldsymbol{\sigma}_1$, $\boldsymbol{\sigma}_2$ and $\boldsymbol{\sigma}_3$ (*none* of which here functions as a GD self-adjoint Hilbert-space operator), is hermitian, self-inverse and traceless, with determinant –1. The matrix $\boldsymbol{\sigma}_3$ is real and diagonal while $\boldsymbol{\sigma}_1$ and $\boldsymbol{\sigma}_2$ are both *off*-diagonal, with $\boldsymbol{\sigma}_1$ real and $\boldsymbol{\sigma}_2$ imaginary--equal to $i\,\boldsymbol{\sigma}_1\boldsymbol{\sigma}_3$ .

*Left* multiplication of $\boldsymbol{a}$ *by* $\boldsymbol{\gamma}$ in Formula (1)—shifting $s$ by the *imaginary* increment $\Delta$--leaves unchanged $Re\ s$, $y$ and $z$. *Right* multiplication of $\boldsymbol{a}$ by $\boldsymbol{\Gamma}^{-1}$ shifts *all* quc Dirac-coordinates, as Formulas (6), (7) and (8) below make explicit, the various right shifts differing importantly.

*Periodicity* of $\boldsymbol{\Psi}$-dependence on *Im s*--the latter thereby a *compact* Dirac coordinate,

$$\boldsymbol{\Psi}\ (s, y, z) = \boldsymbol{\Psi}\ (s \pm 2\pi i, y, z), \tag{4}$$

is implied by Formula (3). The norm (2) is correspondingly to be understood as integration over *any* continuous 2π interval of *Im s*. The constraint (4) specifies *integer* eigenvalues for $\boldsymbol{N}$ –the Dirac-*momentum* self-adjoint *chirality*-operator canonically-conjugate to the Dirac-*coordinate* operator, *Im s*. Compactness of *Im s*, as well as of $\omega$, allows the *pair* of central conserved universe attributes, chirality and electric charge, to play a 'united role' in universe evolution.

The quc-energy self-adjoint Hilbert-space operator, $\boldsymbol{E}$, is the (Dirac-momentum) canonical conjugate of $\tau\ Re\ s$--that we shall call *'quc local time'*. A von Neumann *quc phase space* is that of the *pair*, $\boldsymbol{E}$ *and* $\tau\ Re\ s$. *Arbitrariness* at $\tau = \tau_0$ of $\boldsymbol{\Psi}^\tau\ (\{\boldsymbol{a}_Q\})$ will allow [as in Formula (10)] von-Neumann 'quantum phase space' big-bang-*definition* of hamiltonian-essential *finitely-positive CL*-invariant 'quc masses'--$M\hbar/2\tau c^2$ (*trivial* operators--neither 'Dirac coordinates' nor 'Dirac momenta'). Big-bang-upper-bounded by a huge but finite integer, $M_{max}$, is a set of $M$-labeled adjacent *positive* 'quc-mass' integers.

For each $Q$ there is an $\boldsymbol{E}, \boldsymbol{N}, y, z$ csco with the Dirac-momenta, (continuous) $\boldsymbol{E}$ and (discrete) $\boldsymbol{N}$, canonically conjugate, respectively, to the (continuous) Dirac coordinates $\tau\, Re\, s$ and $Im\, s$. A finite set of quc-mass integers $M$ in the hamiltonian, Formula (13), *collectively* specifies (together with CL-invariant quc-*pair* spatial separations) retarded gravitational potential energy.

Accompanying the *positive* von Neumann quc-integer $M$ are the *paired* quc-chirality-specifying integer-eigenvalue $N$ *and* the quc-electric-charge-specifying integer, $Q$. The latter chiral-electro integer-pair *w*ill below be seen correlated so as to take (collectively) 19 different values. In Formula (10) negative (positive) $Q$ values accompany odd (even) $M$ values.

A continuous *non-compact* $Re\, s_{q},\, y_{q},\, z_{q}$ Dirac-coordinate quintet attaches to each *different* $M, N, Q$ *integer-trio*--the **boldface** index, $\boldsymbol{q}$, designating the latter. CL-algebra commutation-relations distinguish the pair of CL-invariant $\boldsymbol{N}, \boldsymbol{Q}$ central *non*-geometric displacement generators from the *geometric* non-invariant *3-vector Riemann-momentum* components—an exterior-algebra *trio* (see later section on quc kinetic energy) that generates infinitesimal *quc spatial* displacements in some set of 3 mutually-locally-perpendicular directions through a (hyperbolically) curved metricized 3-dimensional Riemannian-base-space.

[Because the foregoing infinitesimal spatial-displacement *directions* are specified in some 'local' Lorentz frame, whereas each geodesic follows a curved path with *parallel transport* of any *locally-perpendicular* axis trio, the later-defined invariant positive self-adjoint quc *kinetic* energy—a ($Q$, *N-independent*) function of *Casimir* geodesic-associated *second* derivatives—is *not* proportional to the inner product with itself of Riemannian 3-vector quc momentum. The later-specified *positive* continuous-spectrum CL-invariant (electro-chiral-*independent*) self-adjoint quc *kinetic energy* is consistent with Riemannian curvature of 'base' 3-space.]

The 6-dimensional Haar measure,

$$d\boldsymbol{a} = ds\, dy\, dz, \tag{5}$$

is CL-invariant, as are *each* of the 3 *separate* factors, $d\,(Re\, s)$, $d\,(Im\, s)$ and $dy\, dz$. [Any 'volume-element' symbol $d\xi$ in (5), with $\xi$ complex, means $d\, Re\, \xi \times d\, Im\, \xi$.]

*Left* CL transformation, $\gamma\boldsymbol{a}$, we have seen to be $s \rightarrow s+\Delta,\ y \rightarrow y,\ z \rightarrow z$, with *imaginary* $\Delta$. *Right* (exterior) transformation, $\boldsymbol{a}\Gamma^{-1}$, by a straightforward computation, leads to

$$z^{\Gamma} = (\Gamma_{22} z - \Gamma_{21}) / (\Gamma_{11} - \Gamma_{12} z), \tag{6}$$

$$y^{\Gamma} = (\Gamma_{11} - \Gamma_{12} z)\, [(\Gamma_{11} - \Gamma_{12} z)\, y - \Gamma_{12}], \tag{7}$$

$$s^{\Gamma} = s + ln\, (\Gamma_{11} - \Gamma_{12} z). \tag{8}$$

Notice how $z$-transformation fails to depend on $y$ or $s$, how $y$-transformation is $s$-independent and *linear* in $y$, how $Re\, s$ and $Im\, s$ transform *independently*--by $\Gamma$, $z$-specified displacements--and how the linear $z$-polynomial, $\Gamma_{11} - \Gamma_{12} z$, appears *repeatedly*.

**Schrödinger-Perpetuated Quc-*Specification* (through an $M$, $N$, $Q$ integer trio)**

CL-invariance of hqu hamiltonian, plus the latter's failure ever to 'create' or 'annihilate' a quc specified by the 3 integers $M, N, Q,$ implies that any quc--identifiable by a boldface index, ***q***, shorthand for the foregoing integer trio--'lives forever' in a 'post-big-bang' universe. 'Evolution' is merely ongoing *quantum-theoretic* quc 'redistribution'. How many different qucs are perpetually being 'Schrödinger-rearranged'? The number is huge although finite.

Extensively discussed by now have been the 7 possible values of the quc-electric-charge-specifying integer *Q*. The following section addresses the quc-chirality-specifying integer, *N*. What about the integer *M*? Formula (10) illustrates how a 'von-Neumann big-bang' correlates *sign* of quc electric charge with oddness or evenness of 'quc-mass' integers *M* within a *finite* set of *adjacent* positive integers, $1, 2 \ldots M_{max}$. Throughout this paper's remainder, a fixed (perpetual) finite set of *different* qucs will be receiving attention.

**'Dirac-Occam' Chirality Hilbert-Space**

Canonically-conjugate to the Dirac coordinate *Im s* is the self-adjoint Dirac-momentum operator whose integer-eigenvalues we denote by the symbol *N* and, in *ħ/2 units*, call *'quc chirality'*. Integer spectrum for chirality ('compact' *Im s*) was implicit in Gelfand's SL (2, c) representation, although GN made no note thereof. [(2)] Our (immediately-below) big-bang restriction [(1)] of *quc* chirality *N* to only 3 integer-values, joining *quc* electric-charge restriction to 7 possibilities, is in 'Dirac-Occam' spirit.

We propose big-bang assignment, to any *charged* quc, *one* of the *N*-value *chirality-trio*, 0 and ±1. To any *charge-less* quc we assign *zero*-chirality. The total number of different *Q*, *N* combos is then 19. *Furthermore,* odd *and* even values of *N*, accompanied by *common M,Q* values, are *never* superposed in hqu Hilbert-space--a 'quantum-statistical' chirality constraint that is remindful of, while distinct from, electric-charge 'super-selection'--long a celebrated feature of *all* quantum theories.

When only *odd* total quc-chirality values are superposed, an aggregate is 'fermionic'; if all superposed totals are *even*, the aggregate is 'bosonic'. *Except* via the quc-specifying subscript, $_q$, the hamiltonian will be seen to *ignore* chirality—to be 'GD-*super*-symmetric'.

For *Q* = 0 qucs, irrelevant to particles but composing galactic-scale gravity-sustained '*dark*-matter' aggregates, we (in Occam spirit) allow *only* the *N*-value 0, while *both* even *and* odd (positive-integer) *M* values. Dirac never contemplated chirality—a notion uncovered by Gelfand's *unitary* SL (2, c) representation.

An appendix of Ref. (1) applies *Q* ≠ 0 'chirality-tripling' (*N* = 0, ±1) to an electron 3-quc Hilbert-vector where *N* = 0 for *both* qucs of a 2-quc electron-mass-determining (approximately) 'Higgs core', while *N* is *either* +1 *or* –1 for a Dirac-doubling 'valence' electron-quc. (Valence-chirality *wave-function* 'Dirac-determines' electron rapidity!) Electron-core qucs have *Q* = +3 *and Q* = –3 while valence-quc has *Q* = –3--*all* electron qucs being 'bright'. [See discussion following Formula (10) concerning *negativity* of (total) electron charge.]

Physics *anti*-symmetry of a *multi*-electron wave function is currently under study. We conjecture that physics 'quantum statistics' accompanies QFT's (huge-age) *disregard* of quantum gravity--allowing Wigner's 10-parameter 'Poincare group' and 'conformal invariance'.

*Valence* quc of muon or tau (*as well as that of electron)* is 'brightly negative' but ('Higgs') muon-*core* has *Q* = +2 and –2, while tau-*core* has *Q* = +1 and –1.

Each of 3 Dirac neutrino or antineutrino quc-*pairs* is a (core-*less*) aggregate of 2 oppositely-charged qucs, *exactly-one* of which has *N* = 0. Representable by *nonzero*-chirality

*charged*-quc *pairs*, in a variety of *charge* pairings, are *all*-three 'QFT-elementary' *vector* bosons. (Mass of reported Higgs-scalar boson suggests a $Q = \pm 1$ pair of *zero*-chirality qucs.)

**Further Remarks**

The duo--$\boldsymbol{E_q}$, $\boldsymbol{N_q}$--of quc-$\boldsymbol{q}$ 'Dirac *momenta'* commutes with the quartet of 'exterior' continuous-spectrum Dirac-*coordinate* operators $y_q$, $z_q$, (whose *canonical-conjugates* this paper *never* addresses). The self-adjoint operator $\boldsymbol{E_q}$, canonically-conjugate to $\tau\, Re\, s_q$, has a *continuous* spectrum *spanning* the real line. [ Remember that 'quc-mass'--neither a Dirac coordinate nor a Dirac momentum--is *discretely* big-bang defined via von Neumann's 'quantum phase space'.]

Each hqu Hilbert vector specifies a *reality* via expectations of quc Dirac-*coordinate* 4-vector current densities of discrete electric charge *and* continuous energy-momentum. (1) Micro-scale particles, macro-scale condensed matter and galactic-scale dark matter are all 'Popper-discernible'. The bachelor-quc reservoir is *not*.

Any 'particle-aggregate' is a micro-scale 'spatial clump' of energy and momentum with integral total charge, integral or half-integral helicity and matching evenness or oddness of total chirality. Meaning for 'particle mass' (including that of electron or photon) can only be *approximate. Low* physics 'uncertainty' for a particle-mass value is rendered possible in a few cases by *hugeness*, at human-lab scale, of present universe age.

Particle-aggregate *physical* ('observable') location is never absolute but, rather, is *relative* to 'nearby' condensed-matter aggregates of *approximately*-zero total charge. Hamiltonian electrodynamics blurs classical charge discreteness for aggregates of condensed-matter, plasma, stars or galaxies. But spanning 'quality space' are the *dimensionalities* of quc energy, $\hbar/2\tau$, of quc chirality, $\hbar/2$, and of quc electric charge, $(\hbar c)^{1/2}$.

Energy, chirality and electric charge span a 3-dimensional 'quality space' which the author believes never to have been *mathematically* recognized despite being in the toolkits of physicists and astronomers. A familiar device to simplify formulas employs units such that $c = \hbar = 1$. We have done so here. Our hamiltonian gravitational and electromagnetic *quc-pair* retarded-potential energies will below be seen proportional, respectively, to dimension-ful $G$ and dimensionless $g^2$—the celebrated *fine-structure constant* whose smallness enabled the spectroscopy that led physicists to quantum theory.

The 'quality 3-vector' component $q_3 = 0, \pm 1, \pm 2,$ *or* $\pm 3$ (7 options), specifies not only *Quc*-$\boldsymbol{q}$'s electric charge, via a factor $g$, but its baryon number—*both* ignored by GN. (2) The Formula (10)-appearing quality-3-vector components $q_1 = 1, 2 \ldots M_{max}$ and $q_2 = 0, \pm 1$ associate respectively to 'quc-mass' and quc-chirality Hilbert-sectors *un*-emphasized in Ref. (2) although definable through von-Neumann and Gelfand math. Our quality-space notation allows electro-chiral appreciation of GD Hilbert-space features that Gelfand did not pursue—big-bang quc-electric-charge *enlarge-ability* and quc-chirality *shrink-ability*.

Although the reality specification proposed by Reference (1) involves positive-lightlike and positive-timelike 4-vector self-adjoint single-quc Dirac-*coordinate* operators, *absent* from hqu are Feynman's *energy-momentum* 4-vectors! In contrast, QFT and S-matrix energy-momentum (boostable) 4-vectors are algebra members of Wigner's 10-parameter 'Poincare group' (*flat* physical 3-space--*without* cosmological status).

*Contraction* of CL-exterior—SL (2, c)--to Euclidean group, in a $\tau \to \infty$ limit where hqu's hamiltonian becomes a *physics* time-displacement generator, promises *approximate* huge-age lab-space hqu status for a QFT based on gravity-neglect with consequent scale invariance and boost-able 'elementary-particle' energy-momentum 4-vectors. A quark's QFT 'color' we are expecting to associate (via geometrical-group contraction) to its *baryonic valence*-quc mass.

**Quc Kinetic Energy—Proportional to a Member of Gelfand's *Unirrep* Csco**

The present section returns to the *original* (*Q*-absent) Gelfand Hilbert-space—functions *either* of a Dirac-coordinate sextet ***a*** (called 'regular basis' by Naimark [(2)]) *or* of 2 Dirac-momentum operators—continuous-spectrum ***E*** and discrete-spectrum ***N***--plus 4 continuous noncompact Dirac coordinates, ***y, z***. Physicists may find what follows a challenge, although Wigner's celebrated treatment of the Lie group SU (2) provides a parallel to Gelfand's SL (2, c) 'unirrep'. [(2)]

The *algebra* (*not* a csco and *not* Dirac-momenta) of the 6-parameter semi-simple non-abelian geometric *exterior*-SL $(2, c)_R$ CL-subgroup [(2)] comprises the conserved components of a '6-vector'—a second-rank *antisymmetric* Lorentz tensor. Three of these 6-vector members correspond to quc angular momentum, ***J*** (a 'Wigner 3-vector'), and three to curved-geometrical Riemann-3-space quc momentum, ***K****/τ* --also a 3-vector with *non*-commuting components. Each exterior-algebra member is representable by a self-adjoint operator on the quc Hilbert space. In the Dirac-coordinate (GN-'regular' *s, y, z*) basis each of these 6 Lorentz-algebra members linearly and homogeneously superposes *first* (partial) derivatives. [(2)]

Two (invariant) CL *Casimirs,* commuting with each other and with *all* 8 of the CL algebra members, are the ('ordinary 3-vector') operator inner product ***K·J*** and the *difference*, ***K·K – J·J***, of two such products—both these Casimirs being homogeneous in Dirac-coordinate-basis (partial) *second*-derivatives. [(2)] Neither of the foregoing forms is positive, but Ref. (2) displays algebraic equivalence to *another* pair of *invariant* self-adjoint operators, one of which has (positive-negative) *integral* eigenvalues while its companion enjoys a *continuous positive* spectrum. Denoting (as did GN) the former by the symbol *m* and the latter by the symbol *ρ*, the algebraic relations are

$$\boldsymbol{K\cdot J} = (\rho/2)\,(m/2) \text{ and } \boldsymbol{K\cdot K - J\cdot J} = (\rho/2)^2 - (m/2)^2 + 1. \tag{9}$$

The positive *continuous*-spectrum, chiral-electro algebra *disregarding*, hqu *single-quc-**q*** kinetic-energy operator we *postulate* to be $\rho_q/2\tau$. (Following therefrom will be 'hamiltonian scale invariance' when gravity is ignored. See paragraph preceding this paper's conclusion.) We *call* (chirality-correlated) $m_q/2$ 'quc helicity'.

A 4-element Quc-***q*** *unirrep* comprises $\rho_q$, $m_q$ and an operator-*pair* $z_{1q}$--representing quc '*momentum*-direction'. This quartet commutes with $\boldsymbol{E}_q$ and $\boldsymbol{N}_q$; the two integers $m_q$ and $N_q$ are either *both* even or *both* odd.

[The big-bang (von-Neumann) 'phase-space'--that defines the positive quc-mass-integer *M*--allows thinking of a 'conjugate', although *not* 'canonical', relationship between the (mutually-commuting) unirrep (non-conserved, non-Dirac) 'Gelfand-momentum' quartet $\rho_q$, $m_q$, $z_{1q}$ and the (mutually-commuting, 'geometrical') Dirac-*coordinate* quartet $y_q$, $z_q$.]

The GD quc-*coordinate* self-adjoint operator-pair $z_q$ corresponds to *direction* of quc-***q*** movement through curved Riemannian 3-space. [(2)] Failure of quc-***q***'s 'velocity-direction' $z_q$ to commute with this quc's momentum-direction $z_{1q}$ is remindful of Schrödinger's term,

‘zitterbewegung’—describing Dirac’s early *unsuccessful* (although Nobel-prize-winning) effort to ‘relativize’ the electron by a ‘doubling’ of Hilbert space through *electric-charge* sign reversal .

**‘Quc-Mass’ Positive Integers, $0 < M \leq M_{max}$ -- Completing Quc-Specification**

Anticipating a CL-invariant hqu-hamiltonian that ‘Noether-conserves’ 8 aggregate-able quc attributes, we now focus on the ‘one dimensional’ Gelfand subspace spanned by the (*common Q* and *common N*) *canonically-conjugate* pair, $\tau$ *Re s* and $\boldsymbol{E}$, of self-adjoint (Dirac) operators. The Dirac coordinate here is $\tau$ *Re s* (‘quc local time’) while the Dirac momentum is $\boldsymbol{E}$ (quc energy). Von Neumann [4] called such a space a ‘Hilbert phase space’ and, by employing (in concert) *gaussian* functions of the *continuous* spectra of two canonically-conjugate (*non*-commuting) operators, defined a ‘2-dimensional’ *discrete* Hilbert-Dirac ‘phase-space’ (‘imitating’, while avoiding, Dirac’s Hilbert-space-inadmissible ‘delta-function’).

We now illustrate the notion of a $\tau = \tau_0$ *purely*-bachelor-quc wave function in an $s_q$, $y_q$, $z_q$ basis, the $Q$ integer taking 7 possible values and the $Q$, $N$ combo 19. A von-Neumann-imitating, $M$-$N$-defining (and thereby $\boldsymbol{q}$-defining) big-bang, that may require refinement, is

$$\Psi_{\tau_0}(s_q, y_q, z_q) = \Pi_q\, |y_q z_q|^{-1} exp\, \{i\, [M\, Re\, s_q + N\, Im\, s_q] - \tfrac{1}{2}\, [\, (Re\, s_q)^2 + ln^2|y_q| + ln^2|z_q|]\}. \qquad (10)$$

$M$ here may take *any positive-integer* value not exceeding $M_{max}$. A colleague--Jerry Finkelstein--proposes that even (odd) $M$ values accompany positive (negative) $Q$ values. *Cosmologically*, where ‘CPT’ and ‘*anti* dark matter’ lack meaning, there is no parallel to the *equivalence*, in *physics*-status, between ‘particles’ and ‘antiparticles’. The present-universe (where *most* qucs reside in a ‘bachelor reservoir’) contains more electrons than positrons. Our discussion of the charged-lepton trio has employed the familiar supposition of *negative* electron charge. Then (according to Jerry), $M_{max}$ is an *odd* integer.

**‘Quality Space’--Spanning 3 Independent Dimensionalities**

The algebra of the symmetry group CL spans a ‘quality space’ associating to the mysterious ‘threeness’ in the number of *independent* dimensionalities displayed by our universe—a feature that delighted Planck and has long been familiar to astronomers and physicists.

Hqu *permanently* comprises a fixed finite set of $\boldsymbol{D}$ *different* qucs—quc creation or annihilation *never* occurring, merely ‘rearrangement’ as varying *aggregates* dynamically form and dissolve (*no* meaning for ‘vacuum’). The hamiltonian-evolved hqu wave function superposes, for $\tau > \tau_0$, *indefinitely-many* products of normed functions, each product with $\boldsymbol{D}$ factors. The *purely-bachelor*-quc big-bang starting wave function at $\tau = \tau_0$ is a *single* product [Formula (10)?].

**Quc-Pair Retarded-Potential Energy**

The hqu hamiltonian *potential-energy* is a sum over $\tfrac{1}{2}\boldsymbol{D} \times [\boldsymbol{D} - 1]$ quc *pairs* of CL-invariant electromagnetic-gravitational retarded potential energies, $V_{qq'}(\tau)$, $\boldsymbol{q} \neq \boldsymbol{q}'$, whose status resembles that in non-relativistic physics of a (slowly-moving) charged, massive particle pair.

Reference (1) proposes *inverse* potential-proportionality to the retardation-representing factor, $e^{\beta_{qq'}} - 1$. The *positive*-spectrum self-adjoint *geometrical* operator, $\beta_{qq'}$ (that *fails* to depend on the *non*-geometrical indices, $Q$, $Q'$, $N$ and $N'$), when multiplied by $\tau$ is CL-invariant shortest (curved, positive) distance *between* spatial locations of qucs $\boldsymbol{q}$ and $\boldsymbol{q'}$. Elsewhere given is $\beta_{qq'}$ relation to $Re\, s_q$, $Re\, s_{q'}$, $y_q$, $y_{q'}$ and $z_q - z_{q'}$.

Beginning with electromagnetism, Ref. (1) postulates the retarded potential,

$$\tau V^{el}_{qq'}(\tau) = g^2[QQ'/9]\,[e^{\beta qq'}-1]^{-1}. \tag{11}$$

The corresponding CL-invariant while 'more geometrical' (via dependence on the indices, $M$ and $M'$) gravitational retarded potential-energy operator is, in units where $G = 1$,

$$\tau V^{gr}_{qq'}(\tau) = -(M/2\tau)\,(M'/2\tau)\,[e^{\beta qq'}-1]^{-1}. \tag{12}$$

The full ($\boldsymbol{q \neq q'}$) quc-pair retarded-potential operator is the sum, $V^{el}_{qq'}(\tau) + V^{gr}_{qq'}(\tau)$. Notice how (11) and (12) exhibit 'Newton-Coulomb' dependence on $\beta_{qq'}$ for $\beta_{qq'} << 1$ but exponentially-decreasing dependence for $\beta_{qq'} >> 1$. Curvature of 3-space plus retardation leads to this behavior. [(1)]

**Hqu Hamiltonian and Schrödinger Equation**

As the case for Schrödinger in 1927, the hqu hamiltonian sums symmetry-group-invariant kinetic-energy and potential-energy operators that do *not* commute. Hqu $\tau$-increasing dynamics proceeds through a multi-*quc* Schrödinger (first-order) differential equation where, at each post-big-bang age, a CL-invariant although age-dependent hamiltonian generates an infinitesimal wave-function change that prescribes the 'immediately-subsequent' universe wave function. Schrödinger's 1927 equation was similar although without potential-energy retardation, with a diagonalizable hamiltonian and based on a 6-parameter Euclidean group with *flat* 3-space displacements (instead of the 8-parameter CL group with *curved* 3-space displacements).

The CL-invariant age-dependent retarded-potential non-diagonalizable hamiltonian operator is

$$H(\tau) = \boldsymbol{\Sigma}_q\, \rho_q/2\tau + \tfrac{1}{2}\boldsymbol{\Sigma}_{q\neq q'} V^{ret}_{qq'}(\tau), \tag{13}$$

the evolution equation for the SMU state-vector being

$$i\partial\boldsymbol{\Psi}(\tau)/\partial\tau = H(\tau)\boldsymbol{\Psi}(\tau). \tag{14}$$

The *initial* state vector [exemplified by Formula (10)] at $\tau = \tau_0$, of *unaggregated* (single-product, 'bachelor') *qucs*, 'super symmetrically' accords chirality a role in universe history that *precludes* 'anti-galaxies' (where 'anti-dark-matter' would accompany antiparticles).

Notice how, in *absence* of gravitational potential energy, our Schrödinger equation becomes conformally ('scale') invariant—dependent only on age *ratios* and thereby manifesting a QFT feature allowing renormalization. Related may be present accuracy of Maxwell's equations for (expectation-defined) *classical* electromagnetic fields.

**Conclusion**

Here advanced is a theory of all 'things'--packaging 'Mach' cosmology, Gelfand's Hilbert-space *unitary* representation of the Lorentz group and Schrödinger-Dirac hamiltonian quantum dynamics. 'Thing' means any (expectation-specifiable) spatial aggregation of energy and electric-charge carried by 'qucs'--an $M_{max}$-order of magnitude set of *different* quantum-universe constituents, *uncorrelated* at a 'von-Neumann big-bang'.

Riemannian-Hamiltonian quantum dynamics, we suggest, has by now correlated *some* 'low kinetic-energy' qucs to build aggregates that include macro-scale 'conscious' condensed matter. Application of our proposal to particle physics has explained the 3 observed 'generations' of 'elementary' fermions. Explanation of *all* 'elementary-particle' masses is promised. Predicted is *absence* of 'anti dark matter'.

Although there has here been no mention of *number theory*, our 'Occam' guesses have involved the Mersenne primes $2^2–1$ and $2^3–1$. In units where $G = \hbar = c = 1$, the age $\tau_0$ of big bang is 1. We shall be unsurprised if dimensionless $g^2$ and $M_{max}$ turn out related, respectively, to the Mersenne primes $2^7–1$ and $2^{127}–1$. The latter might be the huge integer $M_{max}$--upper-bounding the 'masses' of an un-aggregated 'bachelor-quc reservoir' where, we suppose, most universe energy currently resides.

Early-universe 'Darwinian-bootstrap' evolution of *bright*-quc-*pair* double-helix photon-aggregates stands at the top of the author's personal priority list, although he expects 'himself' no longer to be a 'definable aggregate' when the 'miracle of light' becomes understood by humanity. Understanding may involve the fine-structure-constant mystery.

**Acknowledgements**

This paper arose from decades of conversations with Adi Da Samraj, Bruno Zumino, Korkut Bardakci, Don Lichtenberg, Stanley Mandelstam, Juan Maldacena, Nicolai Reshetikhin, Lawrence Hall, David McGoveran, Mahiko Suzuki and, *especially*, with Francis Low, Murray Gell-Mann, Henry Stapp, David Finkelstein, Eyvind Wichmann, Ramamurti Shankar, Shiing-Shen Chern, Ivan Muzinich *and* Jerry Finkelstein. Remarks from Berkeley Chew, Ling-Lie Chau, Dugald Owen and Frank Chew have been helpful. Pauline Chew and James Curley have assisted manuscript preparation.

Crucial has been *ongoing* advice from Jerry Finkelstein (who suggested cosmological correlation between evenness or oddness of quc mass-integer and *sign* of quc electric charge).